# Functional Approach to the Catalytic Site of the Sarcoplasmic Reticulum $Ca^{2+}$-ATPase: Binding and Hydrolysis of ATP in the Absence of $Ca^{2+}$

Antonio Lax,[1] Fernando Soler,[1] and Francisco Fernández-Belda[1,2]

Isolated sarcoplasmic reticulum vesicles in the presence of $Mg^{2+}$ and absence of $Ca^{2+}$ retain significant ATP hydrolytic activity that can be attributed to the $Ca^{2+}$-ATPase protein. At neutral pH and the presence of 5 mM $Mg^{2+}$, the dependence of the hydrolysis rate on a linear ATP concentration scale can be fitted by a single hyperbolic function. MgATP hydrolysis is inhibited by either free $Mg^{2+}$ or free ATP. The rate of ATP hydrolysis is not perturbed by vanadate, whereas the rate of *p*-nitrophenyl phosphate hydrolysis is not altered by a nonhydrolyzable ATP analog. ATP binding affinity at neutral pH and in a $Ca^{2+}$-free medium is increased by $Mg^{2+}$ but decreased by vanadate when $Mg^{2+}$ is present. It is suggested that MgATP hydrolysis in the absence of $Ca^{2+}$ requires some optimal adjustment of the enzyme cytoplasmic domains. The $Ca^{2+}$-independent activity is operative at basal levels of cytoplasmic $Ca^{2+}$ or when the $Ca^{2+}$ binding transition is impeded.

## INTRODUCTION

The $Ca^{2+}$-ATPase protein from sarcoplasmic reticulum (SR) plays a central role in muscle relaxation by coupling active $Ca^{2+}$ transport inside the intracellular organelle to ATP hydrolysis in the cytoplasm. In analogy to the Post-Albers model proposed for $Na^+$, $K^+$-ATPase, the catalytic and transport mechanism is usually analyzed by considering the existence of two different conformations, enzyme with high-affinity $Ca^{2+}$ bound ($E_1Ca_2$) and $Ca^{2+}$-free enzyme ($E_2$) (de Meis and Vianna, 1979). Moreover, the phosphorylation event is defined as $Ca^{2+}$-dependent, i.e., $E_1Ca_2$ can be phosphorylated by ATP but not by Pi, whereas $E_2$ can be phosphorylated by Pi but not by ATP.

X-ray diffraction data on the three-dimensional enzyme structure have confirmed the existence of the $Ca^{2+}$-dependent states (Toyoshima *et al.*, 2000; Toyoshima and Nomura, 2002). When $Ca^{2+}$ is bound to the enzyme, the nucleotide (N), phosphorylation (P), and activation (A) domains of the cytoplasmic headpiece are widely separated giving rise to a so-called "open conformation." In contrast, $Ca^{2+}$ dissociation produces a "closed conformation" where the cytoplasmic domains gather to form a more compact structure. Transition from the closed to the open conformation occurs when $Ca^{2+}$ binds to the enzyme ($E_2 + 2Ca^{2+} \rightarrow E_1Ca_2$) and is related with saturation of the $Ca^{2+}$ transport sites. However, the approach of cytoplasmic domains, which is essential for $Ca^{2+}$ transport, is triggered by ATP ($E_1Ca_2 + ATP \rightarrow E_1PCa_2 + ADP$).

The conventional reaction scheme outlines a cyclic sequence of partial reactions leading to the transport of two $Ca^{2+}$ ions per ATP hydrolyzed. Steady-state measurements of ATP hydrolysis always provide coupling ratios lower than two (Meltzer and Berman, 1984; Yu and Inesi, 1995) and substitution of ATP by pseudosubstrates is an additional source of intramolecular uncoupling (de Meis and Fíalho de Mello, 1973; Fortea *et al.*, 2000; Rossi *et al.*, 1979). ATP hydrolysis is also known to occur at low rates when SR vesicles are incubated in the

---

[1] Departamento de Bioquímica y Biología Molecular A, Facultad de Veterinaria, Universidad de Murcia, Campus de Espinardo, 30071 Murcia, Spain.
[2] To whom correspondence should be addressed; e-mail: fbelda@um.es.

X

presence of $Mg^{2+}$ but absence of $Ca^{2+}$. The origin of the $Ca^{2+}$-independent activity, also named basal activity, was for a while a matter of controversy since it was unclear whether the activity was linked to the $Ca^{2+}$-ATPase or be-longed to a different protein (Carvalho-Alves and Scofano, 1987; Fernandez et al., 1980; Hasselbach, 1964). This is-sue has never been completely clarified. Nowadays, the $Ca^{2+}$-independent component is usually subtracted from the total hydrolytic activity and ignored.

Standard preparations of SR vesicles or purified $Ca^{2+}$-ATPase protein were used and the effect of ligands that are known to interact at the catalytic site, namely $Mg^{2+}$, 5′-adenylyl imidodiphosphate (AMP-PNP), p-nitrophenyl phosphate (pNPP) and vanadate, was explored in different experimental settings. It was our reasoning that faithful information on the catalytic site/mechanism might be obtained by analyzing ATP binding and hydrolysis in the absence of $Ca^{2+}$. Our data are interpreted in the light of the $Ca^{2+}$-ATPase atomic model and a complementary route in the reaction scheme is proposed to account for the observed hydrolysis in the absence of $Ca^{2+}$-dependent transitions. The relevance of this alternative uncoupled route is stressed.

## MATERIALS AND METHODS

### Materials

[$\gamma$-$^{32}$P]ATP was a radiochemical from PerkinElmer Life Sciences and [$^3$H]glucose was from DuPont/NEN. Analytical reagents and the liquid scintillation cocktail (S4023) were obtained from Sigma. pNPP (disodium salt) and p-nitrophenol were from Roche Molecular Biochem-icals. The $Ca^{2+}$ standard solution Titrisol® was purchased from Merck. Ammonium metavanadate was from Acros Organics. Stock solutions of 1 mM orthovanadate were prepared from ammonium metavanadate by the addition of ultrapure water (milli-Q grade) adjusted to pH 10.0 with NaOH. The solution did not contain significant amounts of decavanadate as judged from the absence of yellowish orange color (Hua et al., 1999). Nitrocellulose filter units with 0.45 $\mu$m pore diameter from Millipore and a Hoefer filtration box from Amersham Biosciences were used in the ATP binding experiments.

### $Ca^{2+}$ in the Media

A given free $Ca^{2+}$ concentration was adjusted by adding appropriate volumes of ethyleneglycol-bis($\beta$-aminoethyl ether)-$N,N,N',N'$-tetraacetic acid (EGTA) and $CaCl_2$ stock solutions, as previously described by Fabiato (1988). The computer program used for calculation took into account the absolute stability constant for the $Ca^{2+}$-EGTA complex (Schwartzenbach et al., 1957), the EGTA protonation equilibria (Blinks et al., 1982), the presence of $Ca^{2+}$ ligands and pH in the medium. From the operational point of view, the absence of $Ca^{2+}$ was established by including 1 mM EGTA with no $CaCl_2$ added.

### Sample Preparation

Fast-twitch skeletal muscle from adult female New Zealand rabbit was the starting material to prepare a microsomal fraction, SR vesicles, as described by Eletr and Inesi (1972). $Ca^{2+}$-ATPase was purified by selective extraction of unwanted proteins with deoxycholate, according to method 2 of Meissner et al. (1973). Microsomal membrane and purified protein were aliquoted, frozen in liquid nitrogen and stored at $-80°C$ until use. Protein concentration was evaluated by the colorimetric procedure of Lowry et al. (1951) using bovine serum albumin as a standard.

### Hydrolysis Rates in the Absence of $Ca^{2+}$ and $K^+$

The hydrolytic activity of SR vesicles or purified ATPase was evaluated at $25°C$ in the presence of a non-rate-limiting ATP-regenerating system. For ATP hydrolysis, the standard reaction medium was 20 mM 4-morpholinepropanesulfonic acid (Mops), pH 7.0, 5 mM $MgCl_2$, 1 mM EGTA, 2 mM phospho(enol)pyruvate (PEP), 6 U/mL pyruvate kinase (PK), and 0.2 mg/mL SR protein or purified ATPase. The release of Pi at given times after ATP addition was evaluated according to the Lanzetta et al. (1979) method. When pNPP was the substrate, the $MgCl_2$ concentration was raised to 20 mM and no regenerating system was included. In this case, aliquots of 1 mL reaction mixture were acid quenched at different times with 0.2 mL of ice-cold 10% trichloroacetic acid. Thereafter, samples were centrifuged at 10,000 rpm and $4°C$ for 5 min and 1 mL of supernatants was recovered and alkalinized by adding 50 $\mu$l of 10 N NaOH (final NaOH concentration was 0.5 N). p-Nitrophenol in the supernatant was evaluated by colorimetric reading at 420 nm, using the molar extinction coefficient $\varepsilon = 1.62 \times 10^4$ $M^{-1}cm^{-1}$ (Fernandez-Belda et al., 2001).

### ATP Hydrolysis in the Presence of $Ca^{2+}$ and $K^+$

Initial rates were measured at $25°C$ in a medium containing 20 mM Mops, pH 7.0, 80 mM KCl, 5 mM $MgCl_2$, 1 mM EGTA, 0.986 mM $CaCl_2$ (10 $\mu$M free $Ca^{2+}$), 2 mM PEP, 6 U/mL PK, 1.5 μM calcimycin (A23187), and 0.2 mg/mL SR protein. The reaction was started by adding a given ATP concentration and the Pi accumulation was measured by the Lanzetta *et al.* (1979) procedure.

### Binding of ATP in a Ca$^{2+}$-Free Medium

ATP specifically bound to the enzyme was evaluated by double-labeling radioactive technique (Champeil and Guillain, 1986). SR vesicles at 0.4 mg/mL were resuspended at 25°C in a medium containing 20 mM Mops, pH 7.0, 1 mM MgCl$_2$, and 1 mM EGTA. Aliquots of 1 mL containing the vesicles were layered onto wet Millipore filters placed in the filtration box. Filters under vacuum were rapidly perfused with 1 mL of medium containing 20 mM Mops, pH 7.0, 1 mM MgCl$_2$, 1 mM EGTA, 1 mM [$^3$H]glucose (at approximately 1000 cpm/nmol) and a given [γ-$^{32}$P]ATP concentration between 1 and 50 μM (at approximately 30,000 cpm/nmol). ATP hydrolysis was negligible during the perfusion time. Filters without any further rinsing were dissolved in the liquid scintillation cocktail and $^{32}$P and $^3$H radioactivity was counted. Blank assays were performed by repeating the procedure with protein-free samples. The filter wet volume was calculated with the aid of [$^3$H]glucose as a tracer and used for the subtraction of unspecific ATP bound. The effect of Mg$^{2+}$ on ATP binding was also evaluated by using 5 or 20 mM MgCl$_2$, in both preincubation and perfusion media. Data points at zero Mg$^{2+}$ were obtained by including 10 mM ethylenediaminetetraacetic acid (EDTA) without addition of MgCl$_2$. The vanadate effect on ATP binding was examined by a similar protocol. The above described preincubation and perfusion media containing 5 mM MgCl$_2$ instead of 1 mM were supplemented with a given vanadate concentration in the range of 1 to 100 μM.

### Data Presentation

Plotted values are the average of at least three independent experiments, performed in duplicate, and showing the standard deviation. Binding isotherms for ATP were fitted to the Hill equation: $Y = Y_{max} [ATP]^n/(K_{0.5}^n + [ATP]^n)$, where $Y$ and $Y_{max}$ are fractional and maximal binding, respectively, $K_{0.5}$ is the concentration giving half-maximal binding and $n$ is the Hill coefficient. The Hill equation was also applied to kinetic measurements according to: $v = V_{max} [ATP]^n/(K_{0.5}^n + [ATP]^n)$. Fitting was carried out by nonlinear regression using SigmaPlot software (Jandel Scientific).

### Molecular Modeling

Data of the Ca$^{2+}$-ATPase structure in the presence (Toyoshima *et al.*, 2000) or absence of Ca$^{2+}$ (Toyoshima and Nomura, 2002) were obtained from the Protein Data Bank under accession code 1EUL and 1IWO, respectively. Handling of atomic coordinates and calculations was performed with the software DS ViewerPro 5.0 from Accelrys Inc. (San Diego, CA).

## RESULTS

The microsomal preparation of SR vesicles exhibited ATP hydrolytic activity when assayed in the presence of Mg$^{2+}$ and absence of Ca$^{2+}$. The initial experiments were performed at 25°C and neutral pH in a medium containing 5 mM Mg$^{2+}$, 1 mM EGTA, and 0.2 mg/mL SR vesicles. The ATP concentration was varied between 1 μM and 1 mM, always below the Mg$^{2+}$ concentration, and a PEP/PK mixture was included to keep the ATP concentration constant. The hydrolysis rate in the presence of 5 mM Mg$^{2+}$ and absence of Ca$^{2+}$ displayed hyperbolic dependence when plotted against a linear scale of ATP concentration (Fig. 1(A)). More importantly, the dependence on ATP was identical when the experiments were repeated using a purified enzyme preparation. A clear difference was seen when the hydrolysis data in the absence or presence of 10 μM free Ca$^{2+}$ were plotted against a logarithmic scale of ATP (Fig. 1(B)). The hydrolysis rate in a Ca$^{2+}$-containing medium displayed a sigmoidal dependence that was clearly different from that obtained in the absence of Ca$^{2+}$. When Ca$^{2+}$ was present, the hydrolytic activity at 1 mM ATP was 10-fold higher.

The Ca$^{2+}$-free medium at neutral pH was also used for studying the effect of Mg$^{2+}$ on hydrolytic activity. These experiments were performed at 25°C by using SR vesicles in the presence of excess EGTA. In different series of experiments, the ATP concentration was fixed at 0.2, 0.5, or 1 mM and the Mg$^{2+}$ concentration was varied between 10 μM and 100 mM. The hydrolysis rate, at a given ATP concentration, displayed a bell-shaped dependence when plotted against Mg$^{2+}$ concentration on a logarithmic scale (Fig. 2). As can be seen the plateau phase was dependent on ATP concentration. Hence, maximal rates increased as the ATP concentration was raised from 0.2 to 0.5 and 1 mM, while the Mg$^{2+}$-induced inhibitory effect was dependent on ATP concentration. The descending limb of the curve was always observed when the Mg$^{2+}$ concentration was raised above that of ATP.

Similar experiments were conducted to explore the dependence of hydrolysis on ATP concentration and the relationship with $Mg^{2+}$. This was done by using SR vesicles in a $Ca^{2+}$-free medium. The $Mg^{2+}$ concentration was fixed at a given value of 0.2, 1, or 20 mM and the ATP concentration ranged from 100 M to 10 mM. A bell-shaped dependence with respect to the logarithm of ATP concentration was patent at 0.2 or 1 mM $Mg^{2+}$ (Fig. 3). The inhibitory effect of ATP was manifested when the nucleotide concentration exceeded that of $Mg^{2+}$. The use of a high $Mg^{2+}$ concentration (20 mM) only showed the ascending limb of the ATP dependence curve.

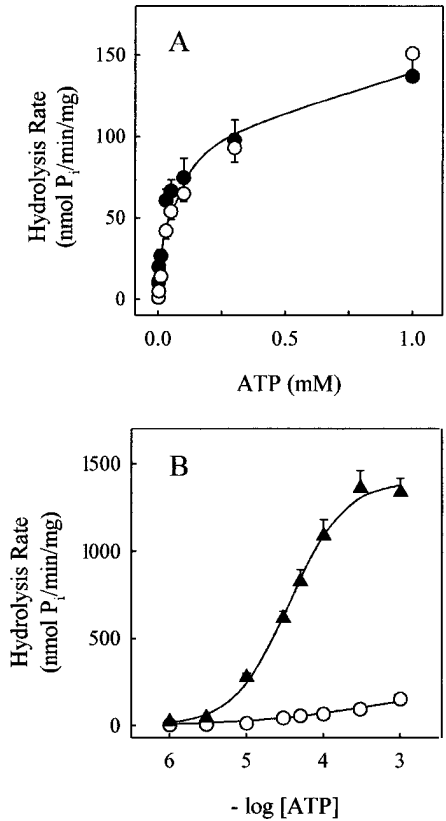

**Fig. 1.** Dependence of hydrolysis rate on ATP concentration measured at 25°C. (A) The experiments were performed in a $Ca^{2+}$-free medium containing 20 mM Mops, pH 7.0, 5 mM MgCl2, 1 mM EGTA, 2 mM PEP, 6 U/mL PK, and 0.2 mg/mL SR vesicles (○) or purified ATPase (●). (B) The experiments were performed in a 10 M free $Ca^{2+}$ medium containing 20 mM Mops, pH 7.0, 80 mM KCl, 5 mM MgCl2, 1 mM EGTA, 0.986 mM CaCl2, 2 mM PEP, 6 U/mL PK, 1.5 M A23187, and 0.2 mg/mL SR vesicles (▲). Data on ATP hydrolysis in the absence of $Ca^{2+}$ plotted in (A) are also shown (○). The hydrolytic reaction was initiated by adding a given ATP concentration and measured by colorimetric reading of Pi.

Additional catalytic properties were revealed by measuring hydrolytic activity in the presence of certain reagents. These experiments were performed at neutral pH in a $Ca^{2+}$-free medium and using SR vesicles. $Mg^{2+}$ was fixed at 5 mM, the ATP-regenerating system was present and the ATP concentration was varied. The dependence of the ATP hydrolysis rate on ATP concentration, when plotted on a logarithmic scale, showed the same profile whether the assays were performed in the absence or presence of 0.1 mM vanadate (Fig. 4). In contrast, the dependence on ATP concentration was shifted to higher concentrations when 0.1 mM AMP-PNP instead of vanadate was included.

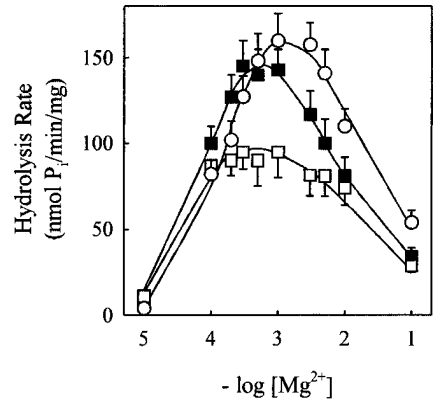

**Fig. 2.** Dependence of ATP hydrolysis rate on $Mg^{2+}$ measured at different ATP concentrations and absence of $Ca^{2+}$. The experimental conditions were 25°C and a medium containing 20 mM Mops, pH 7.0, a given MgCl2 concentration, 1 mM EGTA, 2 mM PEP, 6 U/mL PK, and 0.2 mg/mL SR vesicles. The experiments were started by adding 0.2 (□), 0.5 (■), or 1 mM ATP (○).

The effect of vanadate or AMP-PNP on the rate of pNPP hydrolysis was also evaluated using the same approach. For this purpose, SR vesicles were preequilibrated at neutral pH and in a Ca$^{2+}$-free medium. The Mg$^{2+}$ concentration was 20 mM and pNPP was varied between 1 and 20 mM. The rate of pNPP hydrolysis increased as the pNPP concentration, on a logarithmic scale, was raised (Fig. 5). Interestingly, the hydrolytic profile was the same when the experiments were performed in the presence of 0.1 mM AMP-PNP. However, when the dependence on pNPP was studied in the presence of 1 μM vanadate, in-stead of AMP-PNP, the hydrolytic profile was shifted to higher substrate concentrations.

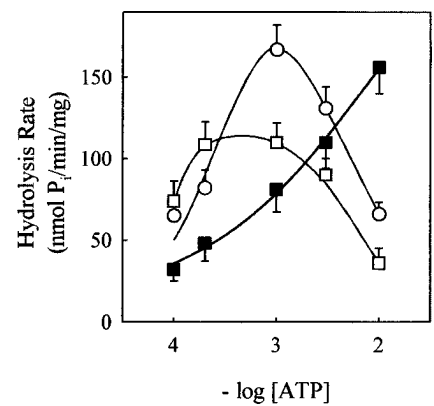

**Fig. 3.** Dependence of ATP hydrolysis rate on ATP measured at different $Mg^{2+}$ concentrations and absence of $Ca^{2+}$. The experiments were performed at 25°C in a medium containing 20 mM Mops, pH 7.0, 1 mM EGTA, 2 mM PEP, 6 U/mL PK, 0.2 mg/mL SR vesicles, and 0.2 (□), 1 (○) or 20 mM MgCl2 (■). The ATP concentration at each fixed $Mg^{2+}$ was varied between 0.1 mM and 10 mM.

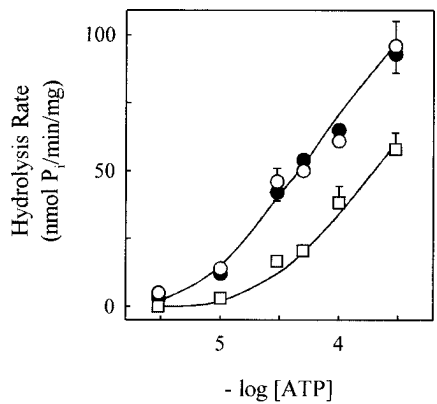

**Fig. 4.** ATPase activity in the absence of Ca$^{2+}$ as a function of ATP concentration: the effect of vanadate or AMP-PNP. Rates of ATP hydrolysis were measured at 25°C in a medium containing 20 mM Mops, pH 7.0, 5 mM MgCl$_2$, 1 mM EGTA, 2 mM PEP, 6 U/mL PK, and 0.2 mg/mL SR vesicles (●). In some experiments, 0.1 mM vanadate (○) or 0.1 mM AMP-PNP (□) was included in the reaction medium before the addition of ATP.

The distinct effect of vanadate on hydrolysis of energy-donor substrates was clearly observed in concentration-dependent experiments. When the standard Ca$^{2+}$-free medium containing the ATP-regenerating system was used and the substrate was 10 μM ATP, the addition of increasing vanadate concentrations up to 100 μM did not alter the hydrolysis rate of SR vesicles (Fig. 6). Nonetheless, the hydrolysis rate was highly sensitive to vanadate when the phosphorylating substrate was 10 mM pNPP. As a reference, the activity in the presence of pNPP was completely inhibited by 10 μM vanadate.

The effect of Mg$^{2+}$ on ATP binding at neutral pH was studied by nucleotide binding experiments under virtually equilibrium conditions. To this end, SR vesicles were preequilibrated in the absence of Ca$^{2+}$ with a given Mg$^{2+}$ concentration. Samples were then rapidly perfused with a medium containing different [γ-$^{32}$P]ATP concentrations, without altering the Mg$^{2+}$ equilibrium, and radioactivity associated with the vesicles was counted. The apparent binding affinity for ATP was clearly diminished when the experiments were performed in the presence of excess EDTA to remove any contaminating Mg$^{2+}$ (Fig. 7) Binding affinity increased when the Mg$^{2+}$ concentration was raised to 1 mM and was even higher at 5 mM. A further increase of Mg$^{2+}$ at 20 mM produced a slight decrease in ATP binding affinity.

ATP binding at the catalytic site was also studied with respect to the Pi-analog vanadate. In these experiments,

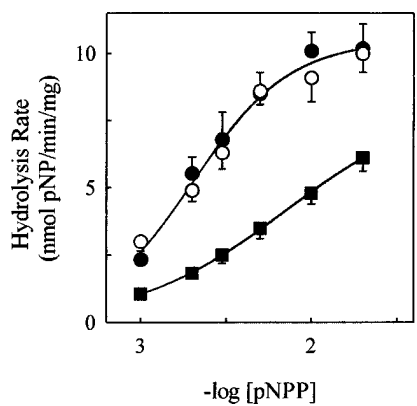

**Fig. 5.** Hydrolytic activity in the absence of Ca$^{2+}$ as a function of pNPP concentration: the effect of vanadate or AMP-PNP. Rates of pNPP hydrolysis were measured at 25°C in a medium containing 20 mM Mops, pH 7.0, 20 mM MgCl$_2$, 1 mM EGTA, 0.2 mg/mL SR vesicles, and a given pNPP concentration (●). The effect of 1 μM vanadate (■) or 0.1 mM AMP-PNP (○) was evaluated by including the corresponding compound in the reaction medium.

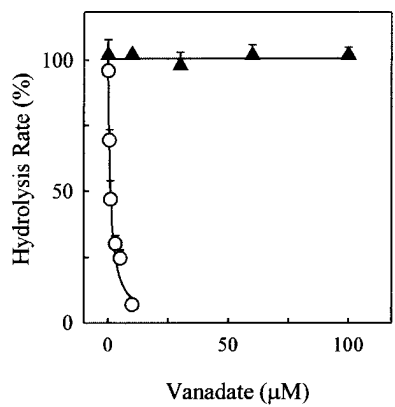

**Fig. 6.** Dependence of hydrolytic activity on vanadate concentration measured in the absence of Ca$^{2+}$: effect of the phosphate-donor substrate. Hydrolysis rates were measured at 25°C in a medium containing: 20 mM Mops, pH 7.0, 5 mM MgCl$_2$, 1 mM EGTA, 2 mM PEP, 6 U/mL PK, 0.2 mg/mL SR vesicles, and 10 μM ATP (▲) or 20 mM Mops, pH 7.0, 20 mM MgCl$_2$, 1 mM EGTA, 0.2 mg/mL SR vesicles, and 10 mM pNPP (○). The reaction medium was supplemented with a given vanadate concentration when indicated.

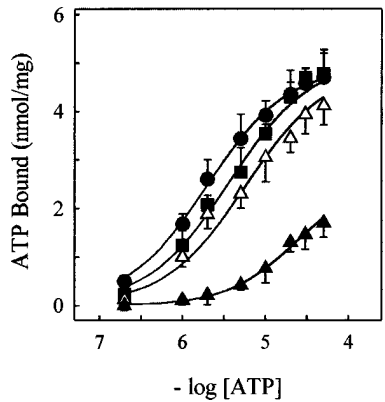

**Fig. 7.** Dependence of ATP binding to the enzyme on ATP concentration: effect of $Mg^{2+}$. One milliliter aliquots containing SR vesicles equilibrated at 25°C in the absence of $Ca^{2+}$ were layered on Millipore filters and subjected to rapid manual perfusion with 1 mL of medium containing radioactive ATP. The preincubation medium was 20 mM Mops, pH 7.0, 1 mM $MgCl_2$, 1 mM EGTA, and 0.4 mg/mL SR vesicles. The perfusion medium was 20 mM Mops, pH 7.0, 1 mM $MgCl_2$, 1 mM EGTA, 1 mM [$^3$H]glucose and a given [$\gamma$-$^{32}$P]ATP concentration of between 0.2 and 50 $\mu$M ($\triangle$). The experiments were repeated by changing $Mg^{2+}$ in both the preincubation and perfusion media to 5 ($\bullet$) or 20 mM ($\blacksquare$). The absence of $Mg^{2+}$ ($\blacktriangle$) was ensured by including 10 mM EDTA in the preincubation and perfusion media without any addition of $MgCl_2$. Radioactivity in the filters, without any rinsing, was used to evaluate specifically bound ATP.

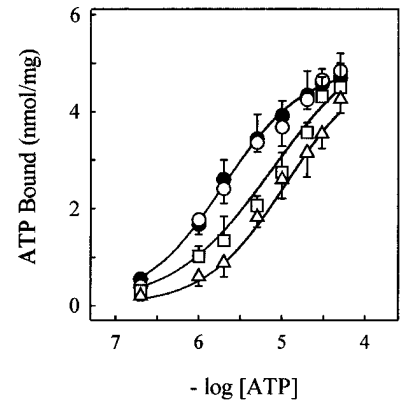

**Fig. 8.** Dependence of ATP binding to the enzyme on ATP concentration: effect of vanadate. Aliquots of preincubation mixture at 25°C and containing 20 mM Mops, pH 7.0, 5 mM $MgCl_2$, 1 mM EGTA, and 0.4 mg/mL SR vesicles were perfused with a medium containing 20 mM Mops, pH 7.0, 5 mM $MgCl_2$, 1 mM EGTA, 1 mM [$^3$H]glucose, and various concentrations of [$\gamma$-$^{32}$P]ATP ($\bullet$). The protocol was repeated by including 1 ($\circ$), 10 ($\square$), or 100 $\mu$M ($\triangle$) vanadate in both preincubation and perfusion media. $^3$H and $^{32}$P in the filters was evaluated by liquid scintillation counting.

$Mg^{2+}$ was fixed at 5 mM and a given vanadate concentration was present in both preincubation and perfusion media. The ATP binding isotherm at neutral pH was unaffected when the perfusion medium containing radioactive ATP was supplemented with 1 $\mu$M vanadate (Fig. 8). However, the presence of 10 $\mu$M vanadate produced a clear decrease in binding affinity. The effect was even more pronounced when the vanadate concentration was raised to 100 $\mu$M.

## DISCUSSION

The rate of ATP hydrolysis in a $Ca^{2+}$-containing medium shows a complex kinetic response with rate-limiting steps usually associated with the $Ca^{2+}$ translocation process (Champeil and Guillain, 1986; Inesi *et al.*, 1982). When $Ca^{2+}$ was absent, the dependence of hydrolysis on ATP concentration followed hyperbolic response when plotted on a linear scale (Fig. 1(A)). On the other hand, the hydrolytic profile in the presence of $Ca^{2+}$ displayed a sigmoidal saturating effect when ATP, on a logarithmic scale, reached the mM level whereas the absence of $Ca^{2+}$ led to lower rates of hydrolysis with no sign of saturation at 1 mM ATP (Figs. 1(B) and 4). The dependence on ATP measured in a $Ca^{2+}$-free medium was the same when microsomal vesicles or purified enzyme were used (Fig. 1(A)) indicating that the hydrolytic activity can be attributed to the $CA^{2+}$-ATPase protein. Therefore, $Ca^{2+}$ binding is not a mandatory requirement for ATP hydrolysis. This observation is supported by previous reports (Carvalho-Alves and Scofano, 1987; Fernandez-Belda *et al.*, 2001; Soler *et al.*, 2002) and does not fit the selective reactivity of $E_1Ca_2$ and $E_2$ with ATP and Pi, respectively, as stated in the coupled reaction cycle.

The $Ca^{2+}$-independent activity has usually been dismissed since it amounts a low percentage of the maximal $Ca^{2+}$-ATPase activity. This is the case when the substrate is ATP (Fig. 1(B)) but may be around 50% or even higher when other pseudosubstrates and/or certain assay conditions are used (Fernandez-Belda *et al.*, 2001; Fortea *et al.*, 2000; Soler *et al.*, 2002). At any rate, a low $Ca^{2+}$-independent activity means that rate-limiting kinetic constants for hydrolysis are significantly lower than those in the presence of $Ca^{2+}$. When the enzyme is in the absence of $Ca^{2+}$ or the $Ca^{2+}$ binding process is blocked all the enzyme population is operating through this route.

The $Ca^{2+}$-independent activity that we measure is expressed by a purified enzyme preparation containing a single electrophoretic band (Fortea *et al.*, 2001) and involved the accumulation of phosphorylated intermediate with an SR $Ca^{2+}$-ATPase-like molecular mass when measured in the presence of UTP and dimethyl sulfoxide (Fortea *et al.*, 2000). Therefore, the participation of $Mg^{2+}$-ATPase activity linked to a minor component of the transverse tubule membrane is unlikely. The Mg$^{2+}$-ATPase is an apparent 105 kDa glycoprotein which is structurally unrelated with the SR Ca$^{2+}$-ATPase, resistant to inactivation by fluorescein isothiocyanate and very sensitive to inactivation by most detergents (Kirley, 1988).

Free Mg$^{2+}$ or free ATP induce enzyme inhibition in the absence of Ca$^{2+}$ (Figs. 2 and 3) if we assume that MgATP is the substrate. An alternative possibility is that the catalytic activity requires the presence of free Mg$^{2+}$ (which is removed by excess ATP) and free ATP (which is removed by excess Mg$^{2+}$). This latter explanation requires additional binding sites for free ligands that are not supported by the crystallographic data.

ATP hydrolysis in the absence of Ca$^{2+}$ is hindered by the presence of the ATP-analog, AMP-PNP, but not by vanadate, whereas pNPP hydrolysis is perturbed by vanadate but not by AMP-PNP (Figs. 4 and 5). In other words, ATP and the Pi-analog vanadate or AMP-PNP and pNPP can be accommodated in the catalytic site at the same time and without mutual interference, but ATP and AMP-PNP or pNPP and vanadate cannot. This is also consistent with the vanadate-sensitivity of pNPP hydrolysis but not ATP hydrolysis in the absence of Ca$^{2+}$ (Fig. 6).

The increase in ATP binding affinity induced by Mg$^{2+}$ (Fig. 7) can be attributed to partial relief of electrostatic repulsion, allowing the -phosphoryl group to closely approach the carboxylate anion of Asp351 and the subsequent nucleophilic attack. This also explains why Mg$^{2+}$ is needed for the phosphorylation of E$_2$ by Pi, the low binding affinity for Pi and the dependence on pH. Likewise, the decreasing effect of vanadate on ATP binding affinity (Fig. 8) can be attributed to the repulsive electrostatic interaction of negative charges at the catalytic site.

It should be noted that $K_m$ for ATP hydrolysis in the absence of Ca$^{2+}$ is 0.5 mM (Fig. 1(A)) whereas ATP binding at 5 mM Mg$^{2+}$ has a $K_d$ around 3 M Fig. 7). This apparent contradiction is due to the fact that $K_m$ is a function of several kinetic constants and therefore, $K_m$ is not a simple measure of the substrate affinity. The lack of correlation between $K_m$ and $K_d$ is greater for the Ca$^{2+}$-independent than for the Ca$^{2+}$-dependent activity. Likewise, ATP hydrolysis in the absence of Ca$^{2+}$ was insensitive to vanadate (Fig. 6) although vanadate may inhibit ATP binding to the enzyme (Fig. 8). A similar paradox was observed when the thapsigargin (TG) effect on ATP binding experiments (DeJesus et al., 1993) was not reflected on ATP hydrolysis measured in a Ca$^{2+}$-free medium (Fortea et al., 2001). It is also known that phospholamban decreases the $K_m$ for SR Ca$^{2+}$-ATPase activation by Ca$^{2+}$ without affecting the Ca$^{2+}$ binding affinity (Lee, 2003).

When arylazido-ATP was the substrate, the dependence on substrate concentration was unaffected by the presence or absence of Ca$^{2+}$ (Oliveira et al., 1988), and the experimental profile was that shown when ATP hydrolysis was measured in the absence of Ca$^{2+}$. A similar dependence was reported with the substrate trinitrophenyl-ATP (Dupont et al., 1985; Watanabe and Inesi, 1982), indicating that an increase in aromaticity of ATP favors substrate utilization through the Ca$^{2+}$-independent route even when Ca$^{2+}$ is present. This is a proof that Ca$^{2+}$-dependent and Ca$^{2+}$-independent are interconvertible hydrolytic activities and both are expression of the Ca$^{2+}$-ATPase protein (Soler et al., 2002).

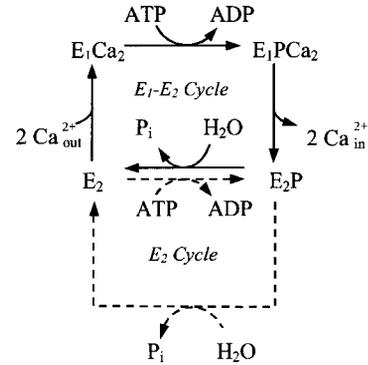

**Fig. 9.** Alternative hydrolytic cycles of SR Ca$^{2+}$-ATPase operating in the forward direction. External Ca$^{2+}$ (2Ca$^{2+}_{out}$) and ATP triggers Ca$^{2+}$ binding: E2 + 2Ca$^{2+}_{out}$ → E$_1$Ca$_2$ and phosphorylation: E$_1$Ca$_2$ + ATP → E$_1$P Ca$_2$ + ADP. Phosphorylation in the presence of Ca$^{2+}$ modifies the packing of transmembrane helices leading to Ca$^{2+}$ entry (2Ca$^{2+}_{in}$) inside the SR vesicles and accumulation of E$_2$P. Hydrolysis of E$_2$P produces Pi release and recovery of E$_2$ with external orientation. This route (solid line) corresponds to a coupled reaction cycle (E$_1$-E$_2$ cycle). When Ca$^{2+}$ is absent in the external medium (E$_2$) or the Ca$^{2+}$ binding transition is blocked, the addition of ATP induces low rates of nonaccumulated E$_2$P + ADP and subsequent release of Pi. This route (dashed line) corresponds to an uncoupled reaction cycle (E$_2$ cycle).

The coupled reaction cycle stands that enzyme phosphorylation and ATP hydrolysis are associated with Ca$^{2+}$ transport provided that the Ca$^{2+}$ binding sites are occupied (E$_1$-E$_2$ cycle in Fig. 9). This coupled cycle becomes uncoupled by the accumulation of free Ca$^{2+}$ inside the SR compartment (Fortea et al., 2000; Meltzer and Berman, 1984; Yu and Inesi, 1995). The present data support the view that ATP hydrolysis also occurs when the Ca$^{2+}$ binding sites are vacant (E$_2$ cycle), although with a different kinetics. This uncoupled cycle does not allow the steady accumulation of phosphoenzyme under standard conditions although partial accumulation can be observed by increasing the substrate (Oliveira et al., 1988) or the medium hydrophobicity (Carvalho-Alves and Scofano, 1987; Soler et al., 2002).

The E$_2$ cycle will be operative under intracellular resting conditions when cytoplasmic free Ca$^{2+}$ is around

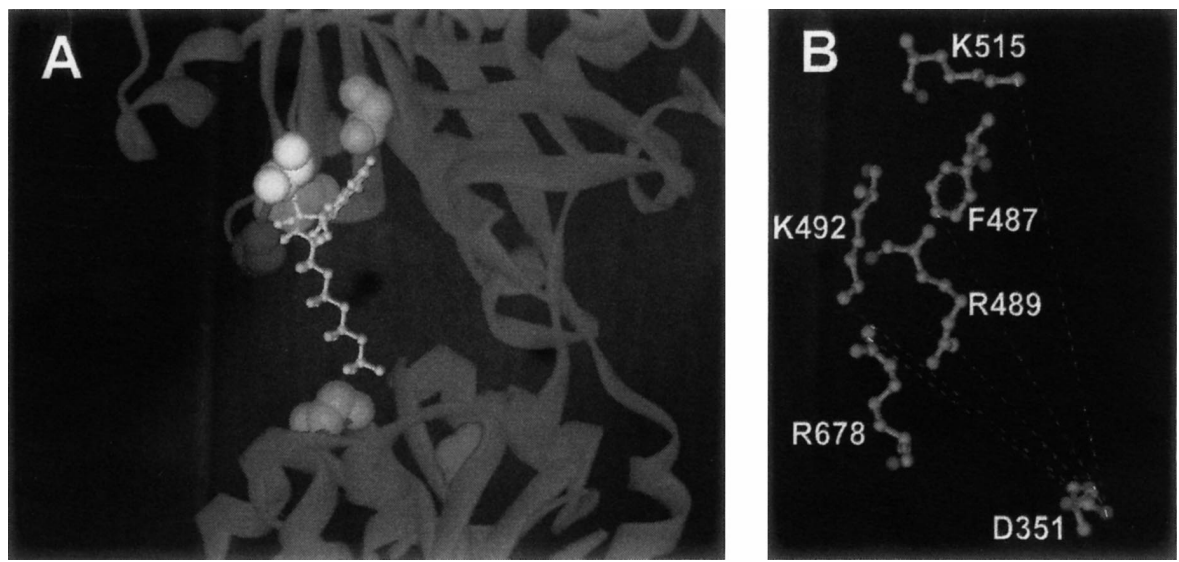

**Fig. 10.** The catalytic site of SR $Ca^{2+}$-ATPase in the $E_2TG$ state. (A) Detail of the enzyme cytoplasmic headpiece showing the catalytic site. The schematic structure is displayed as a deep blue ribbon and relevant residues are shown in space-fill representation. D351, red; F487, light blue; R489, violet; K492, green; K515, pink; and R678, orange. ATP, with the adenine-*anti* conformation and the extended triphosphate moiety, is drawn in yellow. The purine ring is docked in the N domain close to Lys492 and Lys515. For the sake of clarity, the side chain of the selected residues have been removed. (B) Location of selected residues in ball-and-stick representation. C atoms are in *grey*, N in *blue*, and O in *red*. A dashed line is drawn between the side chain extreme of D351 and other residues.

0.1 M. This uncoupled route is also involved in hydrolysis when the $E_2$ conformation is stabilized against $E_1Ca_2$. This latter situation occurs when the enzyme in a $Ca^{2+}$-containing medium is incubated in the presence of TG (Fortea *et al.*, 2001; Sagara and Inesi, 1991) or many other inhibitors of the $Ca^{2+}$-dependent activity, or when the hydrophobic character is increased (Carvalho-Alves and Scofano, 1987; Oliveira *et al.*, 1988; Soler *et al.*, 2002).

ATP binding to the $E_1Ca_2$ conformation cannot directly reach the phosphorylation residue therefore, ATP must induce a conformational change permitting -phosphoryl group to approach Asp351 (Ma *et al.*, 2003; Toyoshima *et al.*, 2000). The putative effect of ATP on the $E_2$ conformation has not been proved but can be envisioned taking advantage of the $E_2TG$ atomic model (Toyoshima and Nomura, 2002). The distance between Lys492 (N domain) and Asp351 (P domain) in $E_2TG$ is 19.3 Å whereas Lys515-Asp351 is 23.8 Å (Table I). It has been suggested that the purine ring of ATP is docked in the N domain close to Lys492 and Lys515. If this is the case, the -phosphoryl group will be situated far away from the phosphorylation center (Fig. 10(A)) and the catalytic utilization of ATP will require some modification of the $E_2$ conformation allowing further approach of the N and P domains. Induced-fit in response to ATP binding in the absence of $Ca^{2+}$ has been reported in several enzymes including adenylate kinase (Sinev *et al.*, 1996).

## ACKNOWLEDGMENTS

This study was funded by grants BMC2002-02474 from Spanish Ministerio de Ciencia y Tecnologia/Fondo Europeo de Desarrollo Regional and PI-22/00756/FS/01 from Fundación Séneca de la Comunidad Autónoma de Murcia, Spain.

**Table I.** Interatomic Distances in the SR $Ca^{2+}$-ATPase Catalytic Site[a]

|  | F487 | R489 | K492 | K515 | R678 |
| --- | --- | --- | --- | --- | --- |
| D351 | 30.5 | 33.9 | 30.9 | 30.2 | 18.0 |
|  | 19.4 | 14.5 | 19.3 | 23.8 | 16.2 |

[a] Structural data were taken from identification codes 1EUL and 1IWO in the Protein Data Bank. Upper row shows distances in $E_1Ca_2$ and lower row corresponds to those in $E_2TG$. Values are expressed in Å and taken between C, N, or O atoms at the side chain extreme of D351 and other relevant residues. Dashed lines in Fig. 10(B) correspond to interatomic distances in $E_2TG$.